\pdfoutput = 1
\documentclass[11pt, dvipsnames]{article}
\usepackage[margin=1in]{geometry}
\usepackage{setspace}
\usepackage[numbers,sort&compress]{natbib}
\let\cite\citep
\usepackage{amsmath}
\usepackage{amssymb}
\usepackage{algorithm}
\usepackage{varwidth}
\usepackage{multirow}
\usepackage{booktabs}
\usepackage{algpseudocode}
\allowdisplaybreaks
\usepackage{hyperref}
\makeatletter
\newcommand{\printfnsymbol}[1]{%
  \textsuperscript{\@fnsymbol{#1}}%
}
\makeatother
\usepackage{graphicx}
\usepackage{caption}
\usepackage{natbib}
\usepackage{subcaption}
\usepackage{color}
\usepackage{bbm}
\usepackage{float}
\usepackage{wrapfig}

\usepackage{amsthm}
\usepackage{dsfont}
\usepackage{mathrsfs}
\usepackage{lmodern}
\usepackage{hyperref}

\usepackage{mathtools}
\usepackage{verbatim}
\usepackage{subcaption}
\usepackage{enumitem}
\usepackage{amsthm}
\usepackage{macros}

\usepackage{booktabs}

\usepackage{xcolor}

\usepackage{appendix}
\title{Hierarchical Generative Agents for \\Simulating Sequential Human Behavior}
\author{%
  Maria G. Mendoza\thanks{Equal contribution.} \quad
  Lucas Waldburger\footnotemark[1] \quad
  Jin Lee \quad
  Shankar Sastry \\
  University of California, Berkeley \\
  {\small \texttt{\{maria\_mendoza, lwaldburger, jinlee\}@berkeley.edu}} \\
  {\small \texttt{sastry@coe.berkeley.edu}}
}
\date{}

\begin{document}

\maketitle
\thispagestyle{empty}
\pagestyle{empty}


\begin{abstract}
Complex cognitive, emotional, and social processes shape human evacuations during natural disasters. Accurate modeling and understanding of human behavior in disasters or emergencies can greatly impact the evacuation process by informing more effective planning and resource allocation. However, collecting human data in these situations is very difficult, and existing computational evacuation models assume rational, homogeneous behavior, leading to unrealistic, overly optimistic predictions. To address this gap, we present a simulation framework of sequential human decision-making during an evacuation scenario, introducing cognitively grounded, persona-driven agents. Our framework models evacuation behavior in a grid-based urban environment that evolves over time, capturing fire and other hazards. Human agents are modeled as personas that make sequential decisions in response to environmental stimuli with cognition structured in three levels: high-level evacuation goals, mid-level route reasoning, and low-level navigation. Decision-making is driven by large language models (LLMs) coupled with a cognitive module and calibrated with empirical human evacuation data. We propose a dynamic, stimulus-driven disaster simulation framework that models human evacuation decision-making using persona-conditioned LLM agents and a cognitive hierarchy. 
\end{abstract}

\section{Introduction}

Inefficient or delayed evacuations during natural disasters result in preventable fatalities. Humans often fail to evacuate promptly due to confusion, misjudgment, or a lack of clear guidance. Disaster response is not purely driven by infrastructure or disaster severity; instead, human behavior is a key determinant of successful evacuation. Decades of research show that humans rarely behave as rational, utility-maximizing agents \cite{choi_AER_2014}, especially in high-stress situations, such as disaster scenarios \cite{bakhshian_evaluating_2023,elhami-khorasani_review_2023}. Empirical evacuation studies document behavioral delays, gathering of belongings, assisting neighbors or family members, strong attachments to property and pets, and the misinterpretation of environmental cues \cite{Xenidis2022-behavioralpattern, verdiere2015human}. These behaviors are further influenced by emotional responses, such as panic or stress, and cognitive biases, such as familiarity seeking in route decision planning or mistrust in alarms \cite{Lindell2012PADM}. Moreover, vulnerable populations, including the elderly, disabled, and those living with dependents, are disproportionately affected, yet are often underrepresented or homogenized in computational models.

Despite extensive study of human behavior in disaster scenarios, existing simulations assume idealized behavioral patterns \cite{bakhshian_evaluating_2023}. These assumptions lead to suboptimal results when incorporated into search-and-rescue technologies, such as drones or other robotic systems, leading to failures when deployed in real-world scenarios \cite{nayyar2019strategies, javdani2015sharedAutonomy}. Consequently, there is a pressing need for simulation frameworks that model heterogeneous social and cognitive decision-making as it evolves over time. 

Recent advances in large language models (LLMs) offer new opportunities to emulate diverse human behavior. However, existing work using heuristics, rule-based systems, or learning-based methods typically focus on one-shot output. As a result, they fail to capture the sequential, time-varying nature of evacuation behavior where decisions unfold over minutes, hours, or days, influenced by both social context and dynamic environmental conditions.

In this work, we introduce a simulation framework for modeling human behavior during disaster evacuation using persona-conditioned LLM agents embedded in a dynamic urban environment.\footnote{Code is available at \href{https://www.github.com/lucaswaldburger/hierarchical_LLM_agents}{github.com/lucaswaldburger/hierarchical\_LLM\_agents}.} The framework enables realistic, long-horizon evacuation modeling by coupling cognitively informed LLM decision-making with environments driven by external stimuli such as hazard progression and limited observability. Beyond behavior modeling, the framework supports applications including emergency resource allocation, evacuation strategy evaluation, alarm system design, urban planning, worst-case scenario analysis, and the testing of human–robot interaction strategies for rescue robotics.  Our contributions are as follows:
\begin{enumerate}
    \item A persona decision-making framework that incorporates empirical evacuation factors informed by survey data \cite{snopkova_predictors_2025} and traditional Protection Action Decision Model (PADM) frameworks, implemented through LLM agents whose decisions better reflect realistic human tendencies (See \autoref{empirical-behaviors}).
    \item A lightweight, easy-to-deploy simulator that captures agent mobility and sequential decision-making in an urban environment with dynamic external stimuli.
    \item A cognitively inspired modeling strategy that leverages LLM reasoning through API calls and multi-level planning to enable long-horizon simulations of diverse, evolving human behavioral responses.
\end{enumerate}

\section{Related Work}
\subsection{Human Behavior in Disasters}
Human behavior during disasters is irrational, heterogeneous, and difficult to predict, yet it plays a decisive role in the outcome of evacuation and rescue operations \cite{Xenidis2022-behavioralpattern}. Factors such as risk perception, social influence, emotional responses (e.g., panic, stress, hesitation), and resource or physical constraints all shape how and when people choose to act \cite{bakhshian_evaluating_2023}. These behaviors can lead to delayed or inefficient evacuations, sometimes with catastrophic consequences \cite{elhami-khorasani_review_2023}. Accurate simulation of evacuation is therefore essential for planning and technology design, but current modeling approaches and data are limited. 

Some evacuation models assume rational actors or focus narrowly on static route optimization and ignore assumptions such as (1) evolving environments, (2) cognitive and emotional variability in response to group dynamics and external stimulus, and (3) partial observability of evacuees. 
To advance evacuation planning, there is a need for models that capture how dynamic environments and social interactions influence human decision-making and generate a range of possible scenarios. 

\subsection{Traditional Rule-Based Methods}
Agent-Based Modeling (ABM) is a widely used approach for simulating multi-agent systems governed by predefined rules, enabling the study of emergent collective behavior, and has been extensively applied to evacuation modeling \cite{SENANAYAKE2024104705}. In this context, ABMs have been used to represent agent navigation, local interactions, communication, social influence, and deviations from rational behavior such as panic \cite{DBLP:journals/isf/SharmaOSG18, trivedi2018panic}. These models enable quantitative analysis of evacuation outcomes, including congestion, bottlenecks, and evacuation times.
Despite these strengths, most ABM approaches rely on fixed rule-based or parametric decision logic, which limits their ability to capture adaptive, sequential, and context-dependent human decision-making under rapidly evolving disaster conditions \cite{Selain-building}.

\subsection{Data-Driven Methods}
Prior work has employed data-driven approaches, including survey and GPS-based mobility data, to study evacuee decision-making and movement patterns during real-world fire events \cite{wang_patterns_2016, KULIGOWSKI2022105541, FORRISTER2024100729}. These datasets enable high-level analysis of evacuation flows, particularly at the city scale, but individual-level route choices are often unavailable due to privacy constraints. As a result, fine-grained evaluation of evacuation behavior across diverse demographic profiles remains limited.

To mitigate these data limitations, Snopkova et al. introduced a controlled virtual-reality dataset capturing individual evacuation route choices, decision times, and route decisions in intersections \cite{snopkova_predictors_2025}. Complementary work has leveraged mixed-reality environments to collect fine-grained human evacuation behavior, including detailed mobility trajectories and routing decisions under simulated fire conditions \cite{mixed-reality-tsinghua}. While the generalizability of such datasets to real-world environments requires further validation, they provide valuable empirical insight into human decision-making processes during evacuation. However, collecting high-fidelity human behavioral data remains costly and difficult to scale.

\subsection{Hybrid Approaches}
Previous studies have explored how LLMs can empower the agency of ABM. Park et al. explored emergent, interactive human behaviors using LLMs \cite{park_generative_2023}. Chopra et al. proposed LLM archetypes to integrate with ABM for large-scale simulations \cite{chopra_limits_2024}. Dai et al. use a persona-aware vision–language model with chain-of-thought reasoning to generate explainable cyclist safety and comfort assessments from street-view imagery \cite{cyclist-VLA-behavior}. Although LLM-integrated agents are known to diversify human decision-making, their effectiveness in emergencies remains uncertain, as disasters often elicit irrational behaviors that are difficult to observe or ethically study. Chen et al. addressed some of these challenges by combining LLMs, survey data, and reinforcement learning to reduce the mismatch between behavior theory and LLM predictions \cite{chen2025wildfire}, but their framework models evacuation as a binary outcome: deciding to evacuate or not. However, effective disaster response requires understanding not only 
\emph{whether} individuals evacuate, but \emph{when} and \emph{how} 
they do so, as evacuation unfolds sequentially over hours or days, with 
highly variable departure times and congestion effects \cite{wang_patterns_2016}. This 
motivates a long-horizon modeling approach that captures a sequence 
of decisions rather than a single-point choice.

\section{Methods}
\subsection{Simulation Environment}
\begin{wrapfigure}{r}{0.65\linewidth}
    \centering
    \includegraphics[width=0.8\linewidth]{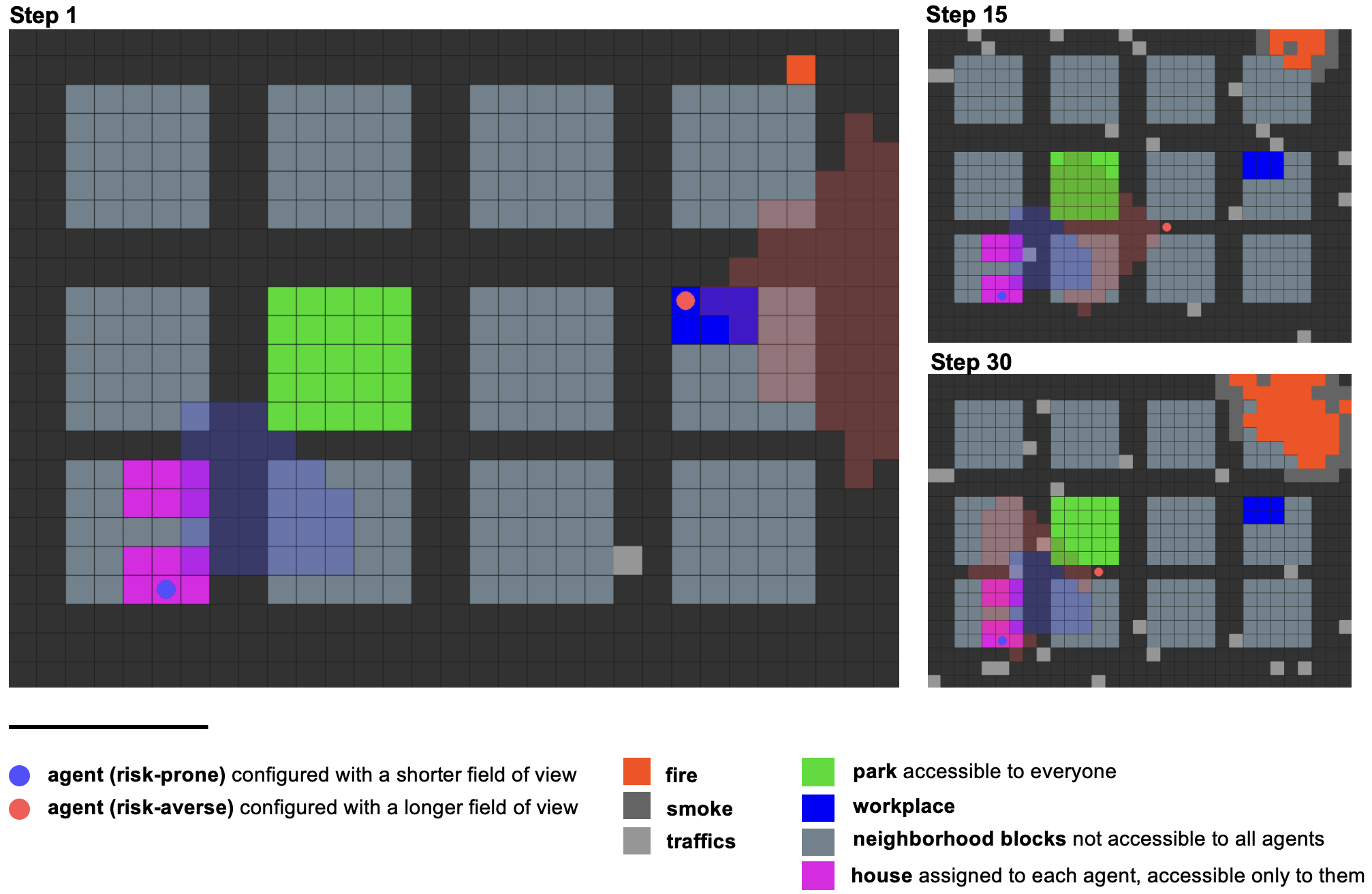}
    \caption{The MiniGrid world with indexed static elements}
    \label{fig:urbanworld}
\end{wrapfigure}

We model an urban environment as a discrete-time, 2D grid world and simulate an evacuation scenario using MiniGrid \cite{MinigridMiniworld23} as shown in Figure \ref{fig:urbanworld}. The world contains static elements such as buildings or homes, roads, and parks. Each element can be accessible or visible to an agent depending on its public or private designation. For example, an agent’s home is only accessible and visible to that specific agent, whereas parks and streets are public and therefore accessible to all agents. We also model spatio-temporal objects such as traffic, fire spreading stochastically, and smoke, which reduces an agent’s visibility and may trigger an emotional response.

We formalize the environment as a finite 2D discrete grid $
\mathcal{X} = \{1, \dots, W-1\} \times \{1, \dots, H-1\} \subset \mathbb{Z}^2,
$
where each cell $x \in \mathcal{X}$, is assigned a semantic type following a labeling function 
$\ell : \mathcal{X} \rightarrow \mathcal{L},$ 
The dynamic environment state at time step $k$ is 
$\mathcal{G}_k = \bigl\langle \mathcal{G}_{\text{static}},\ F_k,\ S_k,\ \mathcal{T}_k,\ e_k \bigr\rangle,$ capturing the static and dynamic elements of the environment, such as fire spread, smoke, traffic, and external stimuli. A detailed formulation is provided in Appendix \ref{environment-def}

\subsection{Personas} \label{persona}
Each agent is modeled as a persona with demographic, cognitive, social, and spatial attributes, which are passed as conditioning variables to the LLM. These traits influence how agents perceive risk, interpret environmental and social cues, experience delays, prioritize goals, and navigate the environment. Persona attributes are informed by empirical evacuation literature and the PADM, capturing heterogeneity in risk perception, trust in alerts, social influence, and pre-evacuation behavior.

\textbf{Demographic Attributes:} Individual characteristics such as name, gender, age, occupation, and workplace location; and  social ties such as friendships, dependents, and propensity to help others.

\textbf{Cognitive and Behavioral Characteristics:} Attributes that shape decision-making during disasters, informed by the PADM. These include risk perception, threat assessment levels, and sensitivity to environmental cues (e.g., alarms, smoke, traffic). In addition, we model susceptibility to social influence (e.g., messages from coworkers or neighbors), likelihood of delaying, ignoring, or acting on alerts. Lastly, we include personal motivations and priorities (e.g., retrieving a pet or belongings) such as housing type (renting vs.\ owning) \cite{FORRISTER2024100729}.

\textbf{Spatial Knowledge and Memory:} Familiar routes, known locations, and place-based preferences that influence navigation.

Persona traits modify every component of the cognitive architecture. They shape the thresholds for interpreting risk, the urgency of updating goals, trust in alarms, willingness to help others, and preference for familiar or unfamiliar routes. These traits are fed into LLM prompts so that each agent responds differently to identical stimuli based on its cognitive and social profile.

Formally, each agent $i \in \mathcal{M}$ is characterized by a persona $\psi^i = \bigl(\psi^i_{\text{att}},\ \psi^i_{\text{cog}},\ \psi^i_{\text{soc}},\ \psi^i_{\text{spat}}\bigr)
$, encoding demographic, social, and spatial attributes. The cognitive parameters $\psi^i_{\text{cog}} = (\rho^i, \tau^i, \theta^i)$ are grounded in the PADM, capturing risk perception, trust in alerts, and threat assessment sensitivity (see Appendix \ref{app-persona} for the full specification).
 
\subsection{Cognitive Architecture}

We designed a cognitive architecture that enables each agent to interact with the environment and make human-centered decisions during disasters. Our approach integrates empirical behavioral data with LLM reasoning to diversify agent responses and model realistic variation across personas. To achieve this, we construct a cognitive module that simulates the decision-making processes most relevant to evacuation scenarios, focusing on the aspects of cognition that meaningfully influence behavior under time pressure and uncertainty. Details of these data sources and how we use the PADM to inform agent behavior are provided in Appendix~\ref{empirical-behaviors}. While individual human responses vary widely, we approximate this diversity by parameterizing personas according to the characteristics described in Section \ref{persona}.

\textbf{Perception:} Each agent has a finite field of view (FOV) representing how far they can observe in the environment. This determines which hazards, obstacles, social cues, and environmental changes they can detect. Perception also determines awareness of other agents, traffic, fire, and smoke.

\textbf{Memory:} The memory module stores key information about the environment and the agent’s past behavior. This includes spatial memory (e.g., familiar routes, accessible areas in the environment), their current and past high-level goals, important decisions previously taken, and key hazard observations. Memory allows agents to maintain continuity in their plan, revise their strategies based on previous failures, and support behaviors such as returning to dependents or avoiding previously dangerous areas.

\textbf{Planning:} As described in Section \ref{multi-level-plan}, we implement a multi-level planning structure that allows agents to make sequential decisions over time. High-level planning governs intent formation in response to external stimuli; mid-level planning determines route choices conditioned on risk perception and persona traits; and low-level planning performs short-horizon navigation based on local observations. This hierarchy reflects how humans adjust plans dynamically as conditions evolve.

\textbf{Reflection:}  
Reflection evaluates urgency, weighs competing priorities, and updates goals. With LLM reasoning, agents consider factors such as whether to trust an alarm, whether to retrieve belongings, or whether to stop and help another agent. Reflection serves as the cognitive ``check'' that allows goals to be revised when conditions meaningfully change.

\textbf{Cognitive Loop and Module Interaction:}
At each timestep, agents operate through a sequential loop:
\textit{Perception} $\rightarrow$ \textit{Memory} $\rightarrow$ \textit{Reflection} $\rightarrow$ \textit{Planning} $\rightarrow$ \textit{Action},
linking environmental observation, state, urgency assessment, hierarchical decision-making, and execution.
    
This cycle repeats continuously, enabling agents to adapt their behavior as the disaster evolves. The full agent state and cognitive module specification are given in Appendix \ref{app-agents}.

\subsection{Multi-Level Planning System} \label{multi-level-plan}
\begin{figure}[h!]
    \centering
    \includegraphics[width=1.0\linewidth]{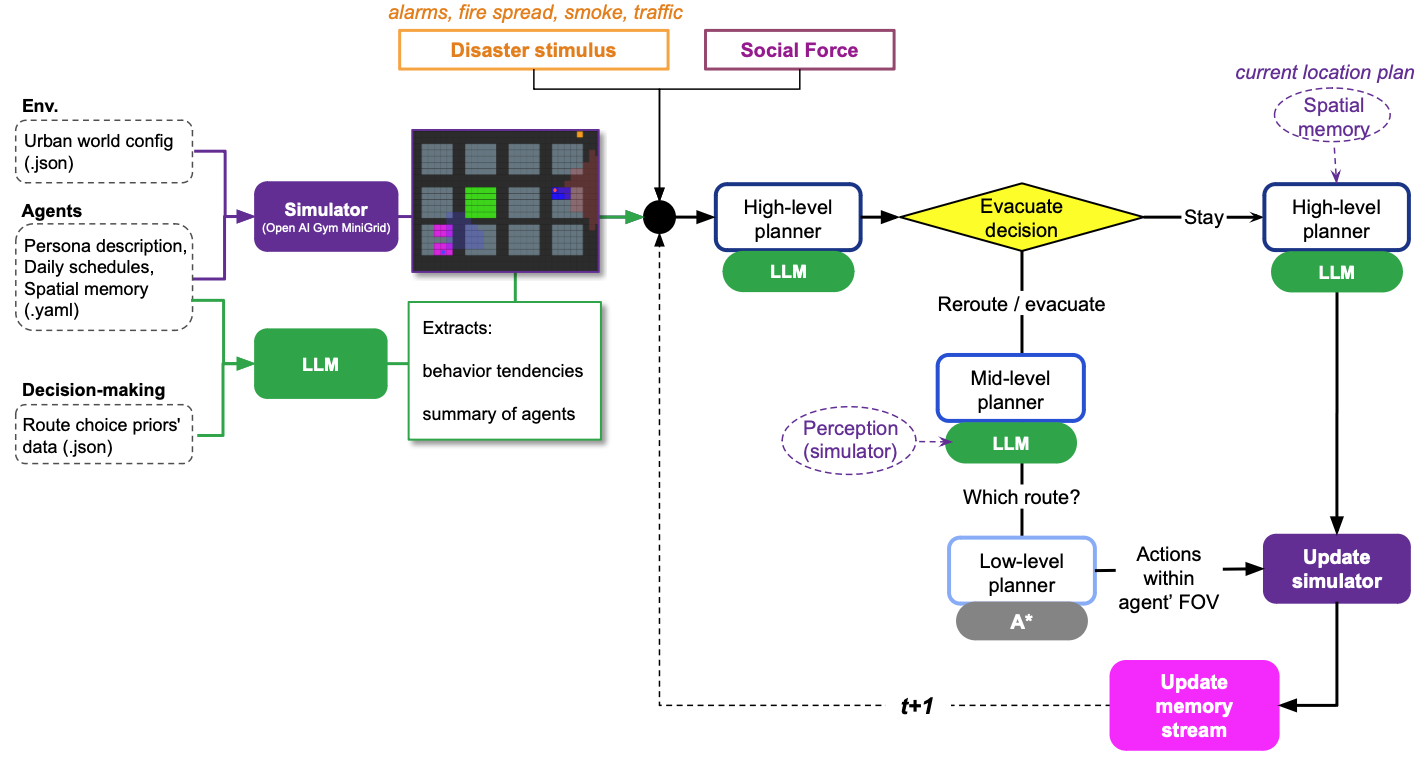}
    \caption{Multi-level decision-making architecture illustrating planner hierarchy, information flow, and event-triggered decision points.}
    \label{fig:multi-level-fig}
\end{figure}

Our framework uses a multi-level planning system, as in Figure \ref{fig:multi-level-fig}, that shows how and when the LLM is prompted, and at which steps the agents make decisions.

The high-level planner engages the LLM when an agent experiences an external stimulus. External stimuli include alarms, official messages, visible smoke thresholds, fire spread, social messages, and environmental changes such as congestion. These triggers prompt high-level reassessment and may override routine plans and are typically broadcast to many or all agents. At this level, the LLM outputs high-level intentions such as ignoring the alert, preparing to evacuate, returning home, or gathering dependents. These decisions are constrained by spatial feasibility, memory, and reflection-based urgency assessments. Alternatively, the LLM may output a decision to evacuate immediately or travel to another location, such as going home to retrieve a pet or prepare belongings. Once a high-level intention is set, the mid-level planner translates it into route decisions, which are then executed step-by-step by the low-level navigator. The agent's policy decomposes as $\pi^i = (\pi^i_H, \pi^i_M, \pi^i_L)$, where $\pi^i_H$ produces a high-level goal $g^i_k \in \mathcal{G}_H = \{\texttt{stay}, \texttt{evacuate}, \ldots\}$, $\pi^i_M$ selects a route $r^i_k \in \mathcal{R}^i_k$ from feasible alternatives, and $\pi^i_L$ executes short-horizon navigation via A* search within the agent's field of view. The full hierarchical planner formulation is in Appendix~\ref{app-multilevel-plan}.

Mid-level planning captures the agent’s route-selection behavior as it moves through the environment, guided by perceived risk, threat level, and traditional PADM principles. We condition the LLM using publicly available behavioral data from \cite{snopkova_predictors_2025}, enabling the agents to generate actions that reflect empirically grounded human decision patterns given their observations and persona (see Section \ref{empirical-behaviors}). At this stage, the LLM outputs a decision on which street or route to follow; for instance, some agents favor familiar paths, others respond more strongly to cues such as visible smoke or congestion, and some exhibit higher risk tolerance that delays protective action. By delegating these decisions to the LLM, the mid-level planner avoids assuming perfectly rational navigation and instead incorporates persona-specific behavioral variability. Once a direction is chosen, the low-level planner executes a short sequence of actions, which is reasonable because over a short horizon, humans tend to follow efficient local movement until new observations prompt reconsideration.

The low-level planner manages short-horizon navigation constrained by the agent’s FOV. While the mid-level planner may designate a general direction (e.g., turn right, proceed along a given street), the agent cannot anticipate obstructions beyond what is locally observable. To capture this uncertainty, movement is restricted to positions within the FOV, and a simple A* search is used to plan only a few steps. This choice serves as a computational approximation of human local navigation behavior, where individuals typically follow the most direct visible path and adapt incrementally to obstacles. The mid and low-level planners aim to improve realism by capturing how decisions evolve with changing observations and computational efficiency by limiting long-horizon pathfinding.

Beyond environmental cues, agents are also influenced by social interactions, which we model through a dedicated layer. The social interaction layer captures interpersonal influences that shape evacuation timing and route choice. These interactions include brief information exchanges, such as a co-worker warning about smoke or encounters with neighbors, where an agent may decide to wait, offer help, or exchange additional information depending on their persona.

\section{Results}

\subsection{Fire Disaster Scenario}

Our simulation consists of $n-$agents in the grid world, where they react to hazards like fire or traffic, and take actions in an evacuation setting. The fire spreads probabilistically based on the number of adjacent fire tiles \cite{Hill_2020}. A tile is set on fire if any of its neighboring cells are burning with probability $p$ = 0.05. Surrounding the fire is smoke, which reduces the FOV size of the agent. 

In addition to the disaster, the agent must navigate through dynamic obstacles, such as traffic. Spatio-temporal density of traffic can be fine-tuned within specific areas in the environment. This is intended to simulate how traffic might be higher in evacuation zones during a disaster condition. Throughout the episode, traffic forces the agent to take detours from their original trajectory, creating congestion that affects route planning and evacuation timing. Together, the fire and traffic produce non-deterministic patterns that test the agent’s decision-making and ability to navigate changing conditions.

\begin{figure}[h!]
    \centering
    \includegraphics[width=0.95\linewidth]{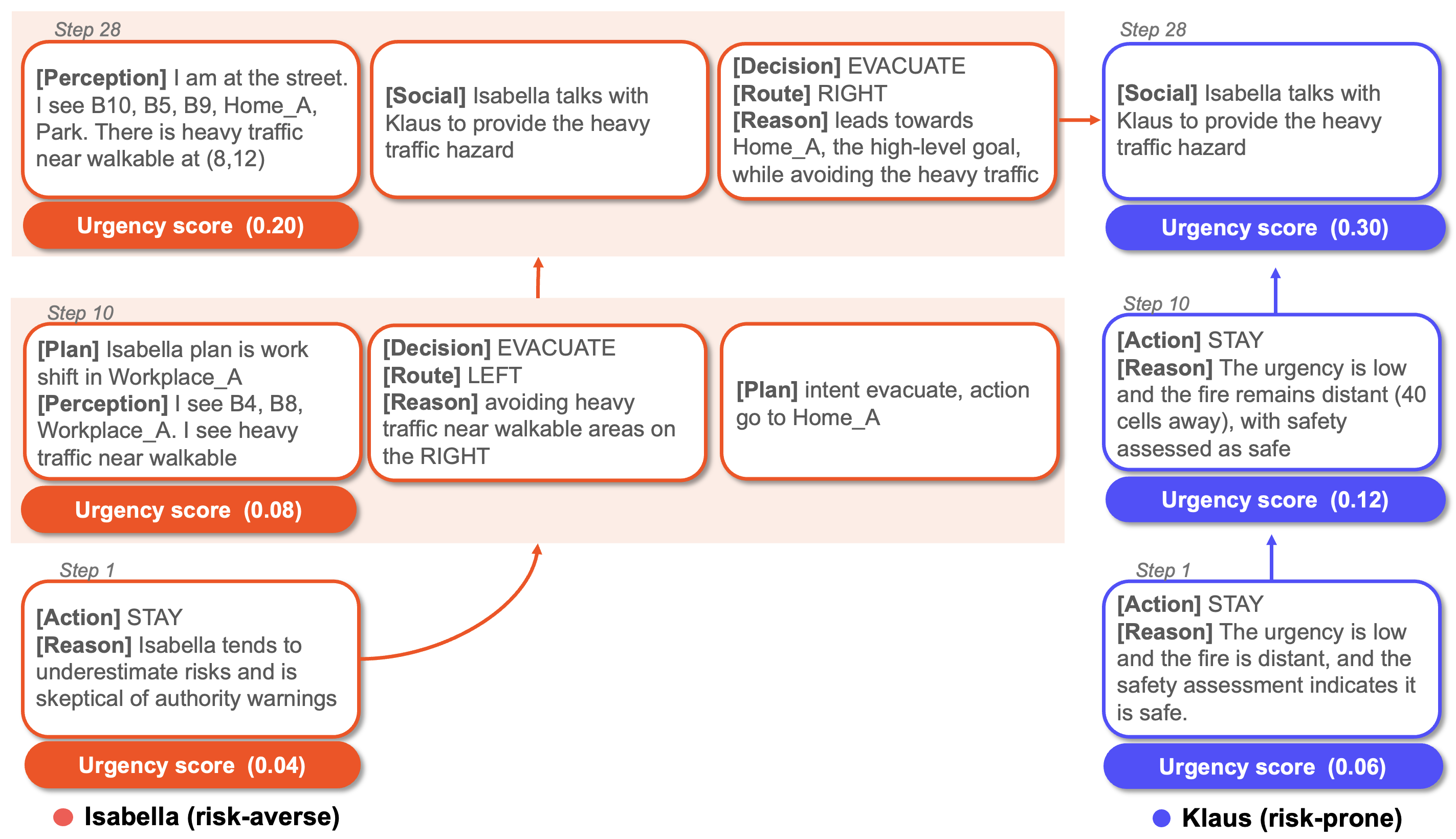}
    \caption{A route-choice tree for Isabella and Klaus based on high (decision), medium (route), and low-level (action) planners.}
    \label{fig:logs}
\end{figure}

Agents begin with different starting contexts (e.g., at home with family or at a workplace) and receive external events of varying severity, eliciting heterogeneous responses. For instance, an agent at work who is risk-prone or exhibits low trust in authorities may choose not to react to an initial text alert indicating the presence of a fire but advising that evacuation is not yet necessary. As hazard urgency increases, such as when smoke is close to the agent, the high-level planner triggers evacuation behavior. Interestingly, persona-based differences emerge naturally among the agents. The risk-averse agent chooses a familiar park as a safe zone, while the risk-prone agent tends to remain in place longer unless the danger becomes unmistakable. At the mid- and low-level, agents dynamically reroute when traffic or fires block their path, exhibiting a preference for familiar streets. Representative outputs are shown in Figure \ref{fig:logs}.

Other agents' behavior outputs shown in Figure \ref{fig:logs} confirms alignment with the PADM analysis workflow (see Figure \ref{fig:Figure5}). The LLM output components, such as the Urgency score, Safety assessment, and Social/environmental cues, are mapped to the sequential steps of the PADM. For example, in Step 28, Isabella's log shows her perception ("heavy traffic near walkable") affecting the Urgency score (0.20), which triggers the decision to evacuate and communicate with Klaus. Isabella's resulting action to move right to avoid the perceived traffic hazard eventually serves as the Low-level goal. This demonstrates how our model captures the PADM steps from threat and risk perceptions to behavioral response. 

\subsection{Behavior Validation} \label{behavior-valid}

We evaluated how well agent characteristics align with observed environmental behavior using a minimal validation scenario. The setup consists of two agents, a risk-prone (blue) and a risk-averse (red) agent, and a static fire hazard (middle red square). The fire is kept static to reduce episode-to-episode variance. Given the hazard, each agent must go to their corresponding home, indicated by pink tiles at the left and right ends of the environment. The risk-averse agent consistently maximizes its distance from the fire by navigating along the boundary of the simulation before returning home. In contrast, the risk-prone agent tolerates closer proximity to the fire compared to the risk-averse agent before rerouting toward a safe location. These results indicate that at a high-level, there is alignment between the agent’s characteristics and their behavior within the environment, while there are minor specifications that need to be optimized, likely through improved prompt engineering.

\begin{figure}[h!]
    \centering
    \includegraphics[width=0.8\linewidth]{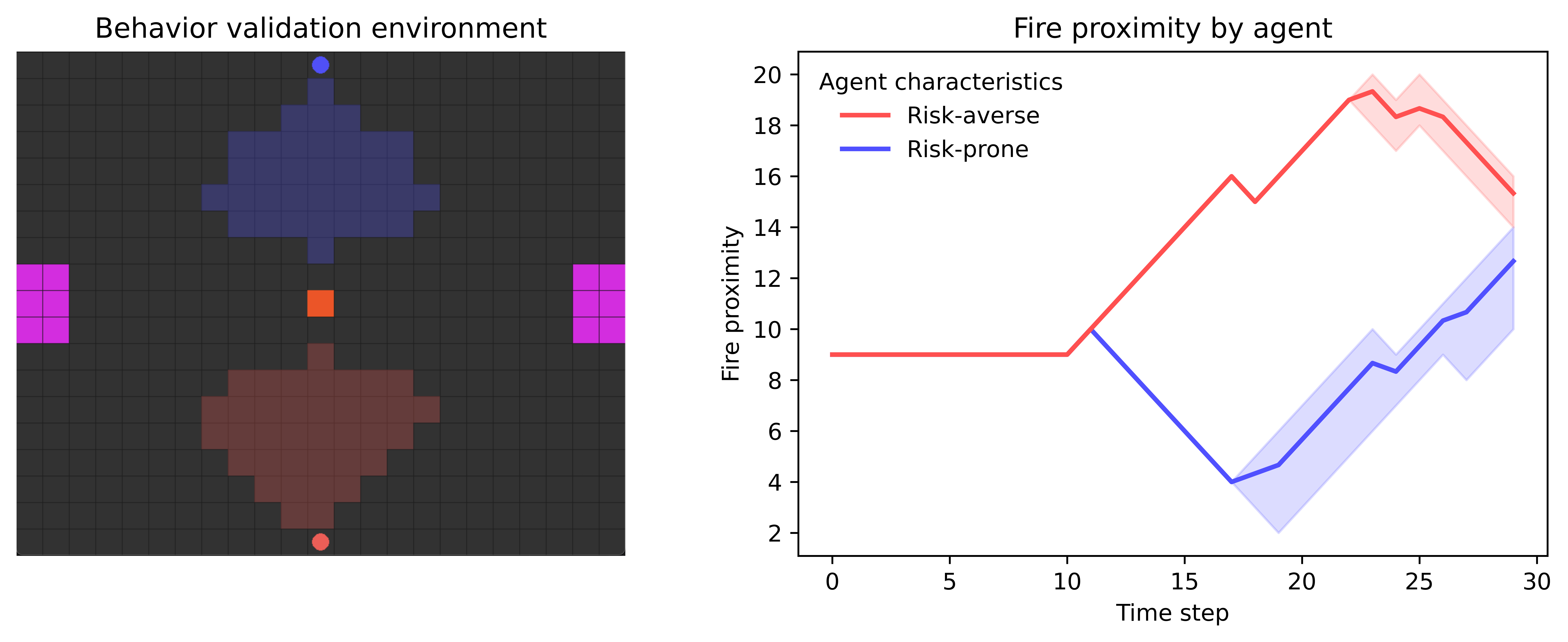}
    \caption{Agent behavior validation. Left: The behavior validation environment where the risk-averse agent (red) and risk-prone agent (blue) must avoid the fire in the middle and go to either of the safe zones (pink). Right: Trends in mean fire proximity as a function of time for agents with different risk profiles.}
    \label{fig:validation}
\end{figure}

\subsection{Scalability}

\begin{figure}
    \centering
    \includegraphics[width=1.0\linewidth]{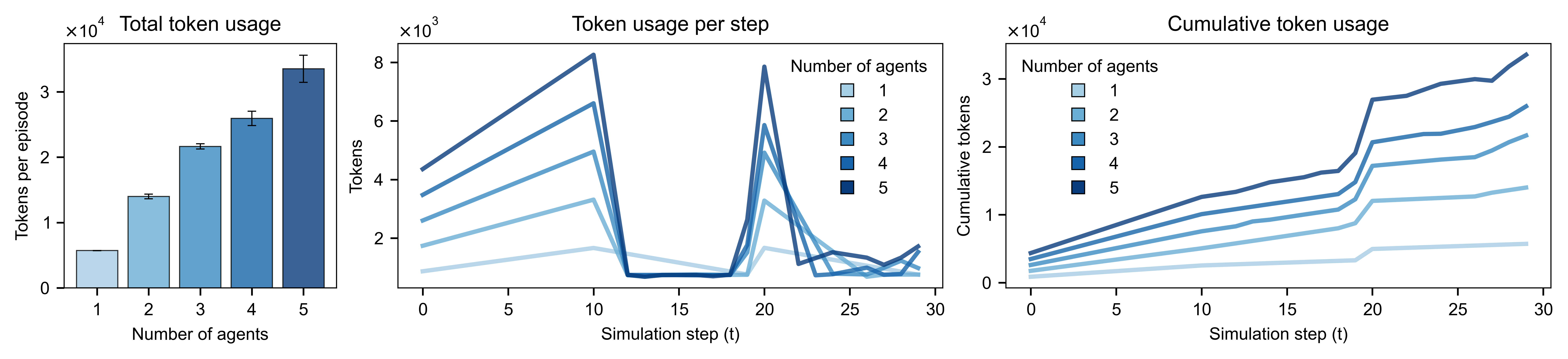}
    \caption{Scalability analysis of the multi-agent system. Left: Distribution of total token usage by the LLM as a function of the number of agents. Middle: Total number of API calls (completions) required per experiment, grouped by agent count. Right: Cumulative token usage over time steps, averaged across runs, illustrating how token consumption grows with the addition of more agents. Each experiment was repeated across 3 episodes.}
    \label{fig:scalability}
\end{figure}

We sought to quantify how well the current implementation scales as a function of agents in order to study emergent behaviors in multi-agent systems in disaster scenarios (Figure \ref{fig:scalability}). Since each agent requires API calls for decision-making, it is important to consider how these computational costs might affect community-level simulation. We ran the fire disaster simulation over three episodes with one to five agents. The number of tokens used increases drastically as more agents are placed in the simulation. The token usage spikes at time steps 10 and 20 when there are three or more agents in the simulation. This likely coincides with time steps where several agents must plan due to an external stimulus and thereby make an API call. We also found that the difference in token usage becomes more volatile as a function of agents since the number of planning steps becomes more variable. These results indicate that the current implementation has limited scalability. Improved scalability can be achieved by using an agent archetype, as in Chopra et al. \cite{chopra_limits_2024} and modifications to the cognitive architecture.

\section{Discussion}
Our framework introduces persona-conditioned LLM agents grounded in behavioral theory and empirical evacuation data for modeling sequential human decision-making during disasters. Through controlled experiments, we demonstrate that persona conditioning 
yields risk-sensitive behavior aligned with the PADM: risk-averse and risk-prone agents exhibit statistically distinguishable hazard proximity profiles (Section~\ref{behavior-valid}), and LLM output components  (urgency scores, safety assessments, social cues) map onto the sequential PADM stages (Figure~\ref{fig:Figure5}). A central challenge for quantitative evaluation of agent-level evacuation models is the absence of ground-truth sequential decision data; existing real-world datasets capture aggregate mobility flows~\cite{wang_patterns_2016} or post-hoc survey responses~\cite{KULIGOWSKI2022105541, FORRISTER2024100729}, not the step-by-step decision traces our model produces. We plan to address this through distributional validation, comparing simulation outputs such as departure-time curves and congestion patterns against documented empirical signatures, and by running agents through the same T-intersection scenarios as human participants in~\cite{snopkova_predictors_2025} to compare route-choice distributions quantitatively. The framework is intentionally lightweight, prioritizing rapid iteration and compatibility with learning-based methods, enabling the study of 
system-level phenomena such as congestion formation, evacuation delays, and bottleneck dynamics.

\section*{Acknowledgments}
This work was supported in part by Provably Correct Design of Adaptive Hybrid Neuro-Symbolic Cyber Physical Systems, DAF Air Force Research Laboratory award number FA8750-23-C-0080.

\section*{References}
\begingroup
\setlength{\bibsep}{0pt}

\renewcommand{\section}[2]{}%
\bibliographystyle{IEEEtran}
\bibliography{CS_294_286}

@article{wang_patterns_2016,
	title = {Patterns and {Limitations} of {Urban} {Human} {Mobility} {Resilience} under the {Influence} of {Multiple} {Types} of {Natural} {Disaster}},
	volume = {11},
	issn = {1932-6203},
	url = {https://dx.plos.org/10.1371/journal.pone.0147299},
	doi = {10.1371/journal.pone.0147299},
	language = {en},
	number = {1},
	urldate = {2025-12-06},
	journal = {PLOS ONE},
	author = {Wang, Qi and Taylor, John E.},
	editor = {Braunstein, Lidia Adriana},
	month = jan,
	year = {2016},
	pages = {e0147299},
}

@article{snopkova_predictors_2025,
	title = {Predictors of evacuation behavior: dataset on respondents’ route choice and web interaction},
	volume = {12},
	issn = {2052-4463},
	shorttitle = {Predictors of evacuation behavior},
	url = {https://www.nature.com/articles/s41597-025-04440-y},
	doi = {10.1038/s41597-025-04440-y},
	language = {en},
	number = {1},
	urldate = {2025-12-06},
	journal = {Scientific Data},
	author = {Snopková, Dajana and Tancoš, Martin and Herman, Lukáš and Juřík, Vojtěch},
	month = jan,
	year = {2025},
	pages = {116},
}

@inproceedings{park_generative_2023,
	address = {San Francisco CA USA},
	title = {Generative {Agents}: {Interactive} {Simulacra} of {Human} {Behavior}},
	isbn = {979-8-4007-0132-0},
	shorttitle = {Generative {Agents}},
	url = {https://dl.acm.org/doi/10.1145/3586183.3606763},
	doi = {10.1145/3586183.3606763},
	language = {en},
	urldate = {2025-12-06},
	booktitle = {Proceedings of the 36th {Annual} {ACM} {Symposium} on {User} {Interface} {Software} and {Technology}},
	publisher = {ACM},
	author = {Park, Joon Sung and O'Brien, Joseph and Cai, Carrie Jun and Morris, Meredith Ringel and Liang, Percy and Bernstein, Michael S.},
	month = oct,
	year = {2023},
	pages = {1--22},
}

@article{elhami-khorasani_review_2023,
	title = {Review of {Research} on {Human} {Behavior} in {Large} {Outdoor} {Fires}},
	volume = {59},
	issn = {0015-2684, 1572-8099},
	url = {https://link.springer.com/10.1007/s10694-023-01388-6},
	doi = {10.1007/s10694-023-01388-6},
	language = {en},
	number = {4},
	urldate = {2025-12-06},
	journal = {Fire Technology},
	author = {Elhami-Khorasani, Negar and Kinateder, Max and Lemiale, Vincent and Manzello, Samuel L. and Marom, Ido and Marquez, Leorey and Suzuki, Sayaka and Theodori, Maria and Wang, Yu and Wong, Stephen D.},
	month = jul,
	year = {2023},
	pages = {1341--1377},
}

@misc{chopra_limits_2024,
	title = {On the limits of agency in agent-based models},
	url = {http://arxiv.org/abs/2409.10568},
	doi = {10.48550/arXiv.2409.10568},
	language = {en},
	urldate = {2025-12-06},
	publisher = {arXiv},
	author = {Chopra, Ayush and Kumar, Shashank and Giray-Kuru, Nurullah and Raskar, Ramesh and Quera-Bofarull, Arnau},
	month = nov,
	year = {2024},
	note = {arXiv:2409.10568},
}

@article{bakhshian_evaluating_2023,
	title = {Evaluating human behaviour during a disaster evacuation process: {A} literature review},
	volume = {10},
	issn = {20957564},
	shorttitle = {Evaluating human behaviour during a disaster evacuation process},
	url = {https://linkinghub.elsevier.com/retrieve/pii/S2095756423000740},
	doi = {10.1016/j.jtte.2023.04.002},
	language = {en},
	number = {4},
	urldate = {2025-12-06},
	journal = {Journal of Traffic and Transportation Engineering (English Edition)},
	author = {Bakhshian, Elnaz and Martinez-Pastor, Beatriz},
	month = aug,
	year = {2023},
	pages = {485--507},
}

@article{chen2025wildfire,
  title        = {From Perceptions to Decisions: Wildfire Evacuation Decision Prediction with Behavioral Theory-informed {LLMs}},
  author       = {Chen, Ruxiao and Wang, Chenguang and Sun, Yuran and Zhao, Xilei and Xu, Susu},
  journal      = {arXiv preprint arXiv:2502.17701},
  year         = {2025},
  url          = {https://arxiv.org/abs/2502.17701}
}

@article{MinigridMiniworld23,
  author       = {Maxime Chevalier-Boisvert and Bolun Dai and Mark Towers and Rodrigo de Lazcano and Lucas Willems and Salem Lahlou and Suman Pal and Pablo Samuel Castro and Jordan Terry},
  title        = {Minigrid \& Miniworld: Modular \& Customizable Reinforcement Learning Environments for Goal-Oriented Tasks},
  journal      = {CoRR},
  volume       = {abs/2306.13831},
  year         = {2023},
}

@article{FORRISTER2024100729,
title = {Analyzing Risk Perception, Evacuation Decision and Delay Time: A Case Study of the 2021 Marshall Fire in Colorado},
journal = {Travel Behaviour and Society},
volume = {35},
pages = {100729},
year = {2024},
issn = {2214-367X},
doi = {https://doi.org/10.1016/j.tbs.2023.100729},
url = {https://www.sciencedirect.com/science/article/pii/S2214367X23001801},
author = {Ana Forrister and Erica D. Kuligowski and Yuran Sun and Xiang Yan and Ruggiero Lovreglio and Thomas J. Cova and Xilei Zhao}
}

@article{KULIGOWSKI2022105541,
title = {Modeling evacuation decisions in the 2019 Kincade fire in California},
journal = {Safety Science},
volume = {146},
pages = {105541},
year = {2022},
issn = {0925-7535},
doi = {https://doi.org/10.1016/j.ssci.2021.105541},
url = {https://www.sciencedirect.com/science/article/pii/S0925753521003842},
author = {Erica D. Kuligowski and Xilei Zhao and Ruggiero Lovreglio and Ningzhe Xu and Kaitai Yang and Aaron Westbury and Daniel Nilsson and Nancy Brown}
}

@article{SENANAYAKE2024104705,
title = {Agent-based simulation for pedestrian evacuation: A systematic literature review},
journal = {International Journal of Disaster Risk Reduction},
volume = {111},
pages = {104705},
year = {2024},
issn = {2212-4209},
doi = {https://doi.org/10.1016/j.ijdrr.2024.104705},
url = {https://www.sciencedirect.com/science/article/pii/S2212420924004679},
author = {Gayani P.D.P. Senanayake and Minh Kieu and Yang Zou and Kim Dirks}
}

@article{Selain-building, author = {Kasereka, Selain K. and Kabwe, Ortega M. and Kinyanta, Maria W.K. and Kasongo, Audrey L. and Ilunga, Godwill W.K. and Muhambya, Katya and Tashev, Tasho and Kyamakya, Kyandoghere}, title = {Enhancing Building Safety: A Brief Review of Agent-Based Modeling for Fire Evacuation Simulation}, year = {2025}, issue_date = {2025}, publisher = {Elsevier Science Publishers B. V.}, address = {NLD}, volume = {257}, number = {C}, issn = {1877-0509}, url = {https://doi.org/10.1016/j.procs.2025.03.086}, doi = {10.1016/j.procs.2025.03.086}, journal = {Procedia Comput. Sci.}, month = jan, pages = {668–675}, numpages = {8} }

@inproceedings{nayyar2019strategies,
title = "Effective Robot Evacuation Strategies in Emergencies",
author = "Mollik Nayyar and Wagner, {Alan R.}",
year = "2019",
month = oct,
doi = "10.1109/RO-MAN46459.2019.8956307",
language = "English (US)",
series = "2019 28th IEEE International Conference on Robot and Human Interactive Communication, RO-MAN 2019",
publisher = "Institute of Electrical and Electronics Engineers Inc.",
booktitle = "2019 28th IEEE International Conference on Robot and Human Interactive Communication, RO-MAN 2019",
address = "United States",
note = "28th IEEE International Conference on Robot and Human Interactive Communication, RO-MAN 2019 ; Conference date: 14-10-2019 Through 18-10-2019",
}

@article{Xenidis2022-behavioralpattern,
title = {Prediction of humans’ behaviors during a disaster: The Behavioral Pattern during Disaster Indicator (BPDI)},
journal = {Safety Science},
volume = {152},
pages = {105773},
year = {2022},
issn = {0925-7535},
doi = {https://doi.org/10.1016/j.ssci.2022.105773},
url = {https://www.sciencedirect.com/science/article/pii/S0925753522001126},
author = {Yiannis Xenidis and Georgia Kaltsidi}
}

@article{verdiere2015human,
  title={Understanding and Simulation of Human Behaviors in Areas Affected by Disasters: From the Observation to the Conception of a Mathematical Model},
  author={Verdière, Nathalie and Cantin, Guillaume and Provitolo, Damienne and Lanza, Valentina and Dubos-Paillard, Edwige and Charrier, Rodolphe and Aziz-Alaoui, Moulay and Bertelle, Cyrille},
  journal={Global Journal of Human-Social Science: H Interdisciplinary},
  volume={15},
  number={10},
  year={2015},
  publisher={Global Journals Inc. (USA)},
  issn={2249-460X},
}

@book{Hill_2020, place={Cambridge}, edition={2}, title={Learning Scientific Programming with Python}, publisher={Cambridge University Press}, author={Hill, Christian}, year={2020}}

@article{choi_AER_2014,
Author = {Choi, Syngjoo and Kariv, Shachar and Müller, Wieland and Silverman, Dan},
Title = {Who Is (More) Rational?},
Journal = {American Economic Review},
Volume = {104},
Number = {6},
Year = {2014},
Month = {June},
Pages = {1518–50},
DOI = {10.1257/aer.104.6.1518},
URL = {https://www.aeaweb.org/articles?id=10.1257/aer.104.6.1518}}

@article{Lindell2012PADM,
  author    = {Michael K. Lindell and Ronald W. Perry},
  title     = {The Protective Action Decision Model: Theoretical Modifications and Additional Evidence},
  journal   = {Risk Analysis},
  year      = {2012},
  volume    = {32},
  number    = {4},
  pages     = {616--632},
  doi       = {10.1111/j.1539-6924.2011.01647.x},
  url       = {https://onlinelibrary.wiley.com/doi/abs/10.1111/j.1539-6924.2011.01647.x},
  eprint    = {https://onlinelibrary.wiley.com/doi/pdf/10.1111/j.1539-6924.2011.01647.x}
}

@inproceedings{javdani2015sharedAutonomy,
  title={Shared Autonomy via Hindsight Optimization},
  author={Javdani, Shervin and Srinivasa, Siddhartha S. and Bagnell, J. Andrew},
  booktitle={Robotics: Science and Systems},
  year={2015}
}

@article{cyclist-VLA-behavior,
  title     = {Persona-aware and Explainable Bikeability Assessment: A Vision-Language Model Approach},
  author    = {Dai, Yilong and Wang, Ziyi and Wang, Chenguang and Zhou, Kexin and Qian, Yiheng and Xu, Susu and Yan, Xiang},
  journal   = {https://arxiv.org/abs/2601.03534},
  year      = {2026}
}

@article{DBLP:journals/isf/SharmaOSG18,
  author       = {Sharad Sharma and
                  Kola Ogunlana and
                  David Scribner and
                  Jock Grynovicki},
  title        = {Modeling human behavior during emergency evacuation using intelligent
                  agents: {A} multi-agent simulation approach},
  journal      = {Inf. Syst. Frontiers},
  volume       = {20},
  number       = {4},
  pages        = {741--757},
  year         = {2018},
  url          = {https://doi.org/10.1007/s10796-017-9791-x},
  doi          = {10.1007/S10796-017-9791-X},
  bibsource    = {dblp computer science bibliography, https://dblp.org}
}

@ARTICLE{trivedi2018panic,
  author={Trivedi, Ashutosh and Rao, Shrisha},
  journal={IEEE Transactions on Computational Social Systems}, 
  title={Agent-Based Modeling of Emergency Evacuations Considering Human Panic Behavior}, 
  year={2018},
  volume={5},
  number={1},
  pages={277-288}}

@article{mixed-reality-tsinghua,
title = {Mixed reality LVC simulation: A new approach to study pedestrian behaviour},
journal = {Building and Environment},
volume = {207},
pages = {108404},
year = {2022},
issn = {0360-1323},
doi = {https://doi.org/10.1016/j.buildenv.2021.108404},
url = {https://www.sciencedirect.com/science/article/pii/S0360132321008015},
author = {Minze Chen and Rui Yang and Zhenxiang Tao and Ping Zhang}
}
\endgroup

\appendix

\newpage
\section{Human Behavior in Disaster}\label{empirical-behaviors}
\begin{figure}[h!]
\centering
\includegraphics[width=0.7\linewidth]{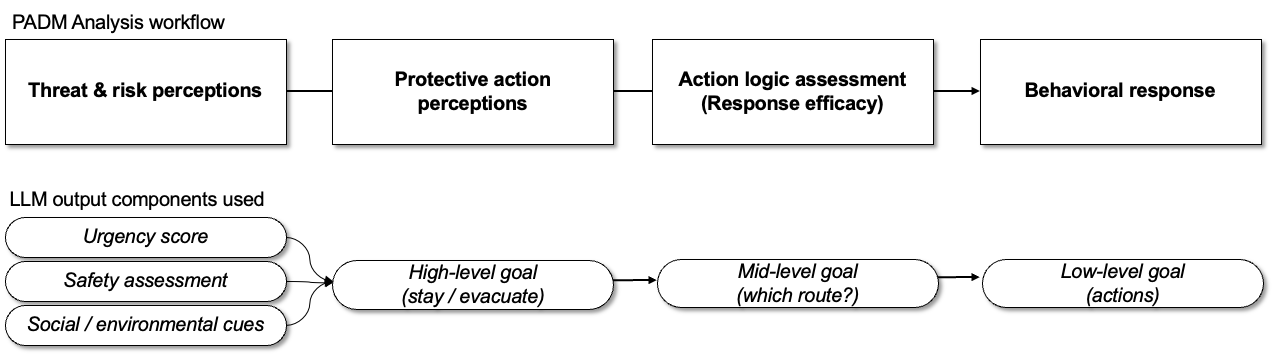}
\caption{Correspondence between the PADM analysis and the LLM output components used.}
\label{fig:Figure5}
\end{figure}

We use the following data to build our framework:
\begin{itemize}
    \item \textbf{Predictors of evacuation behavior \cite{snopkova_predictors_2025}:} provides empirical data on human route-choice decisions during fire evacuations in controlled virtual environments, capturing how environment cues like corridor width, length, and the presence of other transitioning routes influence behavior across different groups. Although the study focuses on human participants navigating simplified building layouts, we leverage this dataset to abstract key decision-making tendencies, rather than replicate individual behavior. In our work, these insights inform how agents with different demographic profiles and risk sensitivities respond to perceived congestion, spatial constraints, and local route preferences. This allows us to ground the LLM reasoning approach in empirically observed patterns while maintaining scalability and generalization to dynamic evacuation scenarios.
    \item \textbf{Human mobility patterns in wildfire \cite{wang_patterns_2016}:} offers the geotagged positional data, capturing human movement patterns before (steady state), during (perturbed state), and after (perturbed state) each wildfire event. This dataset allows for a comparison of mobility characteristics under normal versus disaster conditions. In our project, this dataset influenced how we modeled our disaster scenario, how far people move during the evacuation phase, and the length of the horizon for evacuation. 
    \item \textbf{LLM-based method integrating behavioral theory Protective Action Decision Model (PADM) \cite{chen2025wildfire}:} utilize data collected from local residents via surveys following wildfire events. Here, the PADM, which is a core conceptual framework in disaster psychology, is used to ground the LLM's reasoning. The final output is based on a binary classification (yes or no) for individual evacuation choices, which is simple but shows high predictive accuracy. We refer to their LLM's plan, such as how to integrate the PADM theory with the LLM. Considering PADM, a resident agent's decision-making process focuses on mental states such as risk perception or threat assessment (Figure \ref{fig:Figure5}), especially using prior data \cite{snopkova_predictors_2025} and then selecting relevant variables that influence a resident's perceptions. 
\end{itemize}

\newpage
\section{Formulation}
\subsection{Environment} \label{environment-def}
We model the environment as a finite 2D discrete grid
\[
\mathcal{X} = \{1, \dots, W-1\} \times \{1, \dots, H-1\} \subset \mathbb{Z}^2,
\]
where each cell $x \in \mathcal{X}$ represents a spatial unit of the urban area, and $W, H \in \mathbb{Z}_{>0}$ are fixed dimensions.

Each cell is assigned a semantic land-use type via a static labeling function
\[
\ell : \mathcal{X} \rightarrow \mathcal{L},
\]
where $\mathcal{L}$ is a finite set of mutually exclusive categories. We partition $\mathcal{L}$ into \emph{public} types, accessible to all agents, and \emph{private} types, accessible only to designated agents:
\[
\mathcal{L}_{\text{pub}} = \{\texttt{road},\ \texttt{park},\ \texttt{safe\_zone}\}, \qquad
\mathcal{L}_{\text{priv}} = \{\texttt{building},\ \texttt{workplace},\ \texttt{home}\}.
\]
For each type $\ell \in \mathcal{L}$, we denote the corresponding cell set by $\mathcal{X}_\ell = \ell^{-1}(\ell)$.

\paragraph{Accessibility and visibility.} For a population of agents $\mathcal{M} = \{1, \dots, M\}$, we define two binary predicates for each agent $i \in \mathcal{M}$ and cell $x \in \mathcal{X}$. Let $\text{owns}^i(x) = 1$ if cell $x$ is the private location assigned to agent $i$, and $0$ otherwise.

\begin{itemize}
    \item \textbf{Accessibility:} agent $i$ can physically occupy cell $x$:
    \[
    \mathbf{access}^i(x) = \mathds{1}\bigl[\ell(x) \in \mathcal{L}_{\text{pub}}\bigr]
    \;\vee\;
    \mathds{1}\bigl[\ell(x) \in \mathcal{L}_{\text{priv}} \wedge \text{owns}^i(x) = 1\bigr].
    \]
    \item \textbf{Visibility:} agent $i$ can observe cell $x$:
    \[
    \mathbf{vis}^i(x) = \mathds{1}\bigl[\ell(x) \neq \texttt{building}\bigr]
    \;\vee\;
    \mathds{1}\bigl[\text{owns}^i(x) = 1\bigr].
    \]
\end{itemize}

Roads, parks, and safe zones are both accessible and visible to all agents. Private cells (home, workplace) are accessible and visible only to their assigned agent. Buildings are neither accessible nor visible to any agent.

\paragraph{Hazard dynamics.} For a cell $x \in \mathcal{X}$, let $\mathcal{N}(x) \subseteq \mathcal{X}$ denote the neighborhood of $x$ under a fixed adjacency relation (e.g., 4- or 8-connectivity). At each discrete time step $k \in \{0, 1, \dots\}$, let $F_k \subseteq \mathcal{X}$ denote the set of burning cells. Fire spreads stochastically: each cell $x \notin F_k$ ignites independently according to
\[
\mathbb{P}(x \in F_{k+1} \mid F_k)
= 1 - (1 - p_f)^{|\mathcal{N}(x) \cap F_k|},
\]
where $p_f \in (0,1)$ is the per-neighbor ignition probability. The model assumes fire is permanent: $F_k \subseteq F_{k+1}$. Smoke occupies the traversable cells adjacent to active fire,
\[
S_k = \mathcal{N}(F_k) \cap \{x \in \mathcal{X} : \ell(x) \in \mathcal{L}_{\text{pub}}\},
\]
where $\mathcal{N}(F_k) = \bigcup_{x \in F_k} \mathcal{N}(x)$.
\paragraph{Traffic.} At each time step $k$, traffic is a time-varying occupancy on road cells:
\[
\mathcal{T}_k = \mathcal{T}_k^{\text{exo}} \cup \mathcal{T}_k^{\text{agents}} \;\subseteq\; \mathcal{X}_{\texttt{road}},
\]
where $\mathcal{T}_k^{\text{exo}}$ represents exogenous background vehicles and $\mathcal{T}_k^{\text{agents}}$ captures congestion from evacuating agents. In the current instantiation, $\mathcal{T}_k^{\text{exo}}$ is a time-varying exogenous process parameterized by a spawn rate and density, and $\mathcal{T}_k^{\text{agents}} = \{x \in \mathcal{X}_{\texttt{road}} : |\{j \in \mathcal{M} : c^j_k = x\}| \geq \kappa\}$ for a congestion threshold $\kappa \in \mathbb{Z}_{>0}$.

\paragraph{Environment state.} The static environment collects the structural elements:
\[
\mathcal{G}_{\text{static}} = \bigl\langle \mathcal{X},\ \ell,\ (\mathbf{access}^i)_{i \in \mathcal{M}},\ (\mathbf{vis}^i)_{i \in \mathcal{M}} \bigr\rangle.
\]
The dynamic environment state at time $k$ is:
\[
\mathcal{G}_k = \bigl\langle \mathcal{G}_{\text{static}},\ F_k,\ S_k,\ \mathcal{T}_k,\ e_k \bigr\rangle,
\]
where $e_k$ is the set of external stimuli broadcast at time $k$, such as evacuation orders, alarms, and social warnings.

\subsection{Agents} \label{app-agents}

\subsubsection{Persona} \label{app-persona}

Each agent $i \in \mathcal{M}$ is characterized by a \textbf{persona} $\psi^i$, a fixed attribute vector that conditions all aspects of perception, memory, and decision-making throughout the simulation:
\[
\psi^i = \bigl(\psi^i_{\text{att}},\ \psi^i_{\text{cog}},\ \psi^i_{\text{soc}},\ \psi^i_{\text{spat}}\bigr).
\]

\begin{itemize}
    \item $\psi^i_{\text{att}}$: demographic attributes (age, occupation, housing type, number of dependents).
    \item $\psi^i_{\text{cog}} = (\rho^i, \tau^i, \theta^i)$: cognitive parameters: risk perception $\rho^i \in [0,1]$, trust in alerts $\tau^i \in [0,1]$, and threat assessment sensitivity $\theta^i \in [0,1]$, grounded in the Protective Action Decision Model (PADM).
    \item $\psi^i_{\text{soc}}$: social attributes, including ties to other agents (family, neighbors, coworkers) and susceptibility to social influence $\eta^i \in [0,1]$.
    \item $\psi^i_{\text{spat}}$: spatial knowledge, encoding familiar routes and salient locations within $\mathcal{X}$.
\end{itemize}
\subsection{Observation}

At each time step $k$, agent $i$ does not have access to the full environment state $\mathcal{G}_k$. Instead, it receives a local observation determined by its current position $c^i_k \in \mathcal{X}$ and its field of view.

\paragraph{Field of view.} The observable set of agent $i$ at time $k$ is
\[
\mathcal{V}^i_k = \bigl\{\, x \in \mathcal{X} : \|x - c^i_k\|_\infty \leq \phi^i_k \;\wedge\; \mathbf{vis}^i(x) = 1 \bigr\},
\]
where $\phi^i_k \in \mathbb{Z}_{>0}$ is the effective field-of-view radius at time $k$. Smoke reduces observability: letting $\phi^i_{\text{base}}$ denote the baseline radius (a function of $\psi^i$) and $\Delta\phi > 0$ a fixed reduction,
\[
\phi^i_k = \begin{cases}
\phi^i_{\text{base}} - \Delta\phi & \text{if } c^i_k \in S_k, \\
\phi^i_{\text{base}} & \text{otherwise.}
\end{cases}
\]

\paragraph{Local observation.} The local observation of agent $i$ at time $k$ is the tuple
\[
o^i_k = \Bigl( c^i_k,\; \mathcal{G}_k\big|_{\mathcal{V}^i_k},\; e_k \Bigr),
\]
where $\mathcal{G}_k|_{\mathcal{V}^i_k}$ denotes the restriction of the dynamic environment state to the observable cells $\mathcal{V}^i_k$, and $e_k$ is the set of stimuli broadcast globally at time $k$.

\subsubsection{Memory}

Agent $i$ maintains a memory state $m^i_k$ that is updated at each time step upon receiving a new observation:
\[
m^i_k = \mathrm{Update}(m^i_{k-1},\ o^i_k).
\]
The memory state is a structured record comprising:
\begin{itemize}
    \item \textbf{Spatial memory} $m^i_{\text{spat},k}$: the cumulative map of observed cells $\bigcup_{k' \leq k} \mathcal{V}^i_{k'}$, annotated with cell types, known routes, and locations of salient destinations (home, dependents, safe zones).
    \item \textbf{Hazard history} $m^i_{\text{haz},k}$: the record of observed fire, smoke, and traffic configurations up to time $k$.
    \item \textbf{Goal history} $m^i_{\text{goal},k}$: the sequence of high-level goals pursued by agent $i$ up to time $k$.
    \item \textbf{Decision history} $m^i_{\text{dec},k}$: the sequence of actions taken and their observed outcomes.
\end{itemize}
\subsection{Agent State}

The full state of agent $i$ at time $k$ is:
\[
s^i_k = \bigl(c^i_k,\ o^i_k,\ m^i_k,\ \psi^i\bigr).
\]
Since $\psi^i$ is fixed and $o^i_k$ is determined by $c^i_k$ and $\mathcal{G}_k$, the agent state is fully characterized by the triple $(c^i_k, m^i_k, \psi^i)$. The agent's behavior at each time step is a deterministic or stochastic function of $s^i_k$.

\subsubsection{Cognitive Module} \label{app-multilevel-plan}

The cognitive module governs how agent $i$ maps its current state $s^i_k$ to a decision. It is structured around two components: an \textbf{urgency assessment} and a \textbf{hierarchical planner}.

\paragraph{Urgency.} At each time step $k$, the agent computes an urgency score $u^i_k \in [0,1]$ that quantifies the perceived need for immediate protective action:
\[
u^i_k = f_{\text{cog}}\bigl(o^i_k,\ m^i_k;\ \psi^i_{\text{cog}}\bigr),
\]
where $f_{\text{cog}}$ integrates current observations (proximity to fire and smoke, traffic density, incoming stimuli $e_k$) with memory and persona-conditioned cognitive parameters $(\rho^i, \tau^i, \theta^i)$. Concretely, $u^i_k$ increases with hazard proximity, alarm severity weighted by $\tau^i$, and risk perception $\rho^i$, and is modulated by threat assessment sensitivity $\theta^i$. In our framework, $f_{\text{cog}}$ is implemented via LLM reasoning conditioned on $\psi^i$.

\paragraph{Hierarchical planner.} The agent's decision-making (or policy) $\pi^i$ is decomposed in three levels of planning:
\[
\pi^i = \bigl(\pi^i_H,\ \pi^i_M,\ \pi^i_L\bigr).
\]

\begin{itemize}
    \item \textbf{High-level planner} $\pi^i_H$: activated when $u^i_k \geq \theta^i_H$ or upon receipt of a high-severity stimulus $e_k$. Produces a high-level goal
    \[
    g^i_k = \pi^i_H\bigl(s^i_k,\ u^i_k\bigr) \in \mathcal{G}_H,
    \]
    where $\mathcal{G}_H = \{\texttt{stay},\ \texttt{evacuate},\ \texttt{go\text{-}home},\ \texttt{gather\text{-}dependents},\ \texttt{help\text{-}neighbor}, \dots\}$ is the discrete set of high-level intents.

    \item \textbf{Mid-level planner} $\pi^i_M$: translates the active goal $g^i_k$ into a route decision
    \[
    r^i_k = \pi^i_M\bigl(g^i_k,\ o^i_k,\ m^i_k;\ \psi^i\bigr) \in \mathcal{R}^i_k,
    \]
    where $\mathcal{R}^i_k$ is the set of feasible routes given the agent's current spatial memory and accessibility constraints. Route selection reflects persona-specific behavioral tendencies such as familiarity bias and congestion sensitivity, grounded in empirical evacuation data.

    \item \textbf{Low-level planner} $\pi^i_L$: executes the current route $r^i_k$ over a short horizon $K \in \mathbb{Z}_{>0}$ within the observable set $\mathcal{V}^i_k$, via shortest-path navigation:
    \[
    (a^i_k, \dots, a^i_{k+K}) = \pi^i_L\bigl(r^i_k,\ \mathcal{V}^i_k\bigr),
    \]
    where each $a^i_k \in \mathcal{A} = \{\texttt{N},\ \texttt{S},\ \texttt{E},\ \texttt{W},\ \texttt{stay},\ \texttt{interact}\}$ is a primitive action.
\end{itemize}

\paragraph{Cognitive loop.} At each time step $k$, agent $i$ executes the following sequential cycle:
\[
o^i_k \;\xrightarrow{\text{perceive}}\; m^i_k \;\xrightarrow{\text{assess}}\; u^i_k \;\xrightarrow{\text{plan}}\; (g^i_k, r^i_k) \;\xrightarrow{\text{navigate}}\; a^i_k \;\xrightarrow{\text{act}}\; c^i_{k+1}.
\]
This cycle repeats at every time step, enabling continuous adaptation of behavior as the disaster evolves.

\section{LLM usage statement}
We used \texttt{gpt-4.1-mini} for the agent action calls through the OpenAI API. We also used \texttt{ChatGPT 5.1} to assist in generating the environment JSON files, structuring agent persona files, and grammar/sentence structure assistance in writing.

\newpage

\clearpage

\addtolength{\textheight}{-12cm}   


\end{document}